# Universal superconductivity in FeTe and all-iron-based ferromagnetic superconductor heterostructures

Hee Taek Yi[1,*], Xiong Yao[2], Deepti Jain[1], Ying-Ting Chan[1], An-Hsi Chen[3], Matthew Brahlek[3], Kim Kisslinger[4], Kai Du[1], Myung-Geun Han[4], Yimei Zhu[4], Weida Wu[1], Sang-Wook Cheong[1,5], and Seongshik Oh[1,5,*]

[1] Department of Physics & Astronomy, Rutgers, The State University of New Jersey, Piscataway, New Jersey 08854, United States

[2] Ningbo Institute of Materials Technology and Engineering, Chinese Academy of Sciences, Ningbo 315201, China

[3] Materials Science and Technology, Oak Ridge National Laboratory, Oak Ridge, Tennessee 37831, United States

[4] Center for Functional Nanomaterials, Brookhaven National Laboratory, Upton, New York 11973, United States

[5] Center for Quantum Materials Synthesis and Department of Physics & Astronomy, Rutgers, The State University of New Jersey, Piscataway, New Jersey 08854, United States

*Corresponding authors. Email: taeggy@physics.rutgers.edu; ohsean@physics.rutgers.edu

**Ferromagnetism (FM) and superconductivity (SC) are two of the most famous macroscopic quantum phenomena. However, nature normally does not allow SC and FM to coexist without significant degradation. Here, we introduce the first fully iron-based SC/FM heterostructures, composed of Fe(Te,Se) and $Fe_3GeTe_2$, and show that in this platform strong FM and high-temperature SC robustly coexist. We subsequently discover that chemical proximity effect from neighboring layers can universally drive the otherwise non-superconducting FeTe films into a SC state. This suggests that the ground state of FeTe is so close to the SC state that it could be driven in and out of the SC state with various other perturbations. Altogether, this shows that Fe-Te-based heterostructures provide a unique opportunity to manipulate magnetism, superconductivity and topological physics, paving the way toward new superconducting technologies.**

# 1. Introduction

Despite the strong exclusive natures between superconductivity and ferromagnetism, researchers have long sought the possibility of coupling SC and FM states in the hope of discovering new phenomena and functionalities.[1-8] Particularly over the past decade, topological superconductors (TSC) harboring Majorana states have been attracting significant attention as a platform toward fault-tolerant topological quantum computation.[9-12] One of the promising ways to implement such a TSC state is through spin-triplet superconductivity, which can be stabilized when SC and FM states coexist.[4, 5, 13-15] Among bulk crystals, heavy fermion superconductors such as $UTe_2$ are believed to exhibit coexisting SC and FM states: thus, $UTe_2$ is considered a strong candidate for a TSC.[1, 6, 8, 16, 17] $Sr_2RuO_4$ was once considered a candidate for a TSC exhibiting both SC and FM state, but recent experiments suggest it is not the case.[18] Regardless, all these bulk TSC candidates exhibit relatively low $T_C$ (~1 K) with rare or radioactive elements. An alternative approach is to rely on proximity effects in SC/FM hybrid structures. Although SC/FM proximity effects have been a subject of interest for many decades,[3] there are fundamental limitations in the conventional SC/FM heterostructures in that having an FM layer interfaced with an SC layer inevitably degrades the superconductivity not only through the electronic proximity effect but also through undesirable chemical intermixing.[19-21]

Recently, however, Fe(Te,Se) (FTS) superconductors are rising as a promising TSC platform either on its own or in heterostructures with other telluride layers, such as $Bi_2Te_3$/FeTe, $(Cr,Bi,Sb)_2Te_3$/Fe(Te,Se), FeTe/$Bi_2Te_3$, FeTe/CdTe, and FeTe/MnTe: here and below, 'A/B' implies 'A' grown on top of 'B'.[22-27] These FTS-based heterostructures provide several advantages over conventional proximity platforms in terms of relatively high $T_C$ (> ~10 K), chemical compatibility, and electronic/magnetic tunability. Moreover, one of the surprising observations made in these FTS-based heterostructures is that otherwise non-superconducting FeTe system can become superconducting when it is neighbored by other telluride layers, regardless of them being conducting vs. insulating, topological vs. trivial, or antiferromagnetic vs. ferromagnetic vs. nonmagnetic.[22-28] The origin behind the superconducting FeTe heterostructures remains a mystery to this day.[28] Here, we present the first ferromagnetic superconducting platform composed entirely of Fe-based systems and from this, discover that cationic interdiffusion, or chemical proximity effect, universally leads to superconductivity in FeTe heterostructures.

One of the advantages of the FTS-based SC/FM chalcogenide heterostructures is that both layers share the same (or similar) anions, which allows the effect of anionic interdiffusion to be negligible. Considering that Fe-based compounds are more commonly in the FM state rather than in the SC state, it is not difficult to find Fe-based FM materials. In particular, we note that recently discovered 2D FM materials, $Fe_3GeTe_2$ (FGT) is one of the best Fe-based FM materials that can be mated with the superconducting FTS layer. FGT is a van der Waals material and exhibits strong perpendicular magnetic anisotropy (Ising type) with a Curie temperature near 220 K. There are also reports of Ruderman–Kittel–Kasuya–Yosida (RKKY) type FM in FGT, which may impact the magnetic ordering of adjacent layers. Accordingly, FGT has been one of the popular 2D FM materials used for spintronics research. [29-31] It is also notable that both FGT and FTS are known to be Hund metals, which exhibit correlated electronic properties governed by orbital-selective Hund coupling rather than on-site Coulomb repulsion.[32-35] Accordingly, this FTS/FGT system provides an additional unique opportunity to study how different Hund metals interact with each other and affect the nature of proximity coupling between SC and FM states. Nonetheless, considering that the 6-fold in-plane symmetry of FGT does not match the 4-fold symmetry of the FTS system, it is still questionable whether FTS/FGT heterostructure can grow epitaxially with atomically sharp interfaces.

## 2. Results and discussion

We grew a series of FTS/FGT films on $10 \times 10 \times 0.5$ mm$^3$ Al$_2$O$_3$ (0001) substrates using a custom-built MBE system with a base pressure of $\sim 10^{-10}$ Torr. Before growth, each Al$_2$O$_3$ substrate was cleaned *ex situ* with UV-generated ozone and *in situ* by heating to 750 °C under oxygen pressure of $5 \times 10^{-7}$ Torr. High-purity elemental sources of Fe, Ge, Bi, Se, and Te were thermally evaporated using standard effusion cells. The source fluxes were calibrated *in situ* by quartz crystal micro-balance and *ex-situ* by Rutherford backscattering spectroscopy. Figure 1a shows the schematic SC/FM heterostructure with a sharp streaky RHEED pattern recorded during the growth of each material. 6-fold in-plane symmetry of the FGT layer can be confirmed in the RHEED patterns which exhibit a periodicity of 60 degrees, with an alternating spacing ratio of √3*a* and *a* in two high-symmetry directions. On the other hand, the RHEED pattern of the 4-fold symmetry of the FTS layer on top of FGT shows an alternating spacing ratio of √2*a* and *a* in two high-symmetry directions, with 30 degrees of periodicity. This is consistent with the 12-fold symmetry of FTS films grown on 6-fold symmetric underlying layers of $Bi_2Te_3$ (BT) and MnTe (MT), as observed previously. [23, 24] As shown in Figure 1b, all XRD peaks can be identified with the c-axis (00n) peaks of FTS (squares), FGT (circles), and substrate $Al_2O_3$. The high-angle annular dark-field scanning transmission electron microscopy (HAADF-STEM) image in Figure 2a shows an atomically sharp interface between FTS and FGT with the c-direction lattice constants of 0.62 and 1.64 nm, respectively.

High-quality FTS growth on FGT underlayer is unexpected and surprising. Although we previously observed similarly high-quality epitaxial growth between 4-fold FTS and 6-fold BT as well as 6-fold MT through the hybrid symmetry epitaxy mode, they still required uniaxial lattice matching: their in-plane lattice constants are related by 2:√3 ratio. On the other hand, in the case of FTS vs. FGT with in-plane lattice constants of 3.80 Å (FTS) and 3.99 Å (FGT), they are ~5% off with their original lattice constants and ~10% off when 2:√3 ratio is considered. The discovery that 4-fold FTS can grow on 6-fold FGT even without uniaxial lattice matching implies that the hybrid symmetry epitaxy mode is much more common than we previously anticipated. This could be related to the underlying FGT layer being a van der Waals material with shared Fe ions.

Now that we have identified the structural integrity of the FTS/FGT system, we will discuss their transport properties. In Figure 2b, we present temperature-dependent longitudinal (black) and Hall (red) resistances in the absence of external magnetic fields. Near room temperature, $R_{xy}$ is negligible but as temperature drops below ~220 K, $R_{xy}$ steeply increases and maintains a large value until it plummets to zero, as $R_{xx}$ also goes to zero, at ~13 K. This implies that the FTS/FGT system goes into an FM state below ~220 K, the Curie temperature of the FGT layer, and then fall into the SC state at ~13 K, the superconducting transition temperature of the FTS layer.[31, 33, 36, 37] In Figure 2b, we define and indicate three different superconducting critical temperatures: $T_{onset}$ of 13.5 K, $T_0$ of 9.5 K, and $T_c$ of 12.5 K (when resistance drops to 50% of the normal state resistance), which are very similar to those reported previously for non-magnetic, FTS/Bi$_2$Te$_3$ heterostructures with similar Te/Se composition of the FTS layer.[23] Figure 2c presents Hall resistance vs external magnetic field at various temperatures. It also shows that the coercivity field, $B_c$, keeps increasing as temperature drops below ~220 K until the SC state kicks in below ~13 K. Analysis of the transport properties near $T_c$ (supplementary Figure S1) shows that the system can be well described by the 2D Berezinskii-Kosterlitz-Thouless (BKT) transition.[38-40]

Below $T_c$, even though transport signals all disappear to zero due to superconductivity, we can still probe the ferromagnetic state of the FTS/FGT system using magnetic force microscopy (MFM), as shown in Figure 3a (see supplementary Figure S2, for the complete set of MFM images). The uniform magnetic contrast shown in Figures 3a(I) and (II) represents a downward magnetization state. After the sample is saturated with -1 T of magnetic field (3a(I)), the sample remains in the downward magnetization state up to 0.2 T (3a(II)). As the field further increases, the sample is gradually aligned with the direction of the external field (Figure 3A (III-VI)). Eventually, this leads to the formation of an upward magnetization domain (Figure 3a(VII)). Figure 3a(VIII) shows that this saturated upward magnetization state persists while the magnetic field is reduced to zero. The field dependence of the magnetization signal (black circles in Figure 3b bottom panel) obtained from the MFM data is fully consistent with the anomalous Hall data in Fig. 3b (top panel), measured above the superconducting transition temperature. The persistence of the saturated state in a zero magnetic field with a large coercive field (0.25 T) indicates a strong FM state with uniaxial (perpendicular) magnetic anisotropy of FGT, coexisting with the superconducting state of FTS. The variation in domain population, $\delta f_{rms}$, is estimated by calculating the root-mean-square (RMS) values of MFM signal in each image (red triangles in Fig. 3b bottom panel). The maximal RMS value indicates the largest variation of the magnetic stray field due to the multi-domain state, and the corresponding magnetic field is fully consistent with the coercive field $H_c$, measured with transport in Figure 3b (top panel).

Both superconducting and magnetic properties of the FTS/FGT heterostructures show that there is, if at all, very little degradation of each of their original properties, unlike conventional SC/FM heterostructures, where superconducting $T_c$ always degrades due to interfacial disorders and electronic/magnetic/chemical proximity effects.[19-21] On one hand, this implies that the effect of disorder is minimal in this system and that SC properties of the FTS layer are robust against coupling with the FM order. On the other hand, however, it raises the question of whether there is any proximity coupling between FTS and FGT. To address this question, we synthesized a simpler, Se-free, FeTe/FGT heterostructure, which is supposed to be non-superconducting. It has long been the general consensus that FeTe, the Te-rich endmember of the FTS system, is an antiferromagnetic (AFM) Hund metal rather than a superconductor in its ground state, unlike its superconducting cousin, FeSe.[34, 36] To confirm the baseline properties of FeTe, we first grew FeTe film on a mica substrate, which also has a hexagonal structure like FGT. Its R vs. T curve in Figure. 4a exhibits the characteristic hump feature near 70 K, indicative of the formation of AFM order, without any hint of SC transition: this is consistent with the well-established transport properties of the FeTe system.[36, 41] However, when we grow FeTe on top of the FGT layer in the same growth condition as the FeTe film on mica, all of a sudden, superconductivity arises with a clear transition near 12 K, while a strong FM state develops below ~220 K (see supplementary Figure $S$3 for temperature-dependent Hall effect curves).

The emergence of a superconducting state in otherwise non-superconducting FeTe thin films is surprising. Yet, this is in line with the recent similar observations made in the heterostructures of $(Cr,Bi,Sb)_2Te_3$/FeTe, FeTe/$Bi_2Te_3$, FeTe/CdTe, and FeTe/MnTe.[22-27] When the superconductivity was first observed in FeTe films after a topological insulator (TI), $Bi_2Te_3$ film was deposited on top; it was suspected that the topological surface state (TSS) should be the key factor behind the emerging superconducting state in the FeTe layer. However, subsequent discoveries of similar superconductivity, even in non-topological systems such as FeTe/MnTe and FeTe/CdTe, suggest that TSS is not the critical factor behind the SC state in FeTe. As summarized in Figure. 4b, the only common factor behind all these superconducting FeTe films is that they are interfaced with some telluride layers.[28] Nonetheless, it remains a mystery how neighboring telluride layers can induce superconductivity in the otherwise non-superconducting, antiferromagnetic FeTe films.[28]

When two telluride layers are neighbored by each other, even though the anionic interdiffusion can be ignored, cationic interdiffusion can still occur. In the case of the FeTe/FGT system, both Fe (cation) and Te (anion) interdiffusion can be ignored, but interdiffusion of Ge ions cannot. In order to test the impact of Ge diffusion into the FeTe system, we prepared 20 nm-thick Ge-doped FeTe films on mica substrates while varying Ge doping levels. Surprisingly, when FeTe films are doped with a small amount (0.5 % and 0.75 %) of Ge, clear superconducting transitions showed up near 11 K, whereas 5 % Ge-doped FeTe film displayed insulating behavior without any hint of superconducting transitions, as shown in Figure 4c (see supplementary Figure $S$4a, for the $R_{xx}(T)$ data before normalization). This finding indicates that minute levels of Ge diffusion, beyond detection limits of XRD or STEM, can suppress the AFM ordering and consequently promote superconductivity in the FeTe films. Then, the emerging superconductivity in the FeTe/FGT system is very likely due to a small amount of Ge ions diffusing into the FeTe layer.

According to this observation, the superconducting FeTe films in other telluride heterostructures could also be of similar origin. In order to test this possibility, we carried out similar doping tests on FeTe films using Bi, Mn, Pb, and Cr as dopants, which include the cations used in the reported superconducting FeTe heterostructures.[22-25, 27] Figure 4d shows that all the Bi-, Mn-, Pb, and Cr-doped FeTe films exhibit similar superconducting transition temperatures without any neighboring telluride layers (see supplementary Figure $S$4b, for $R_{xx}(T)$ before normalization). Notably, these five elements - Ge, Bi, Mn, Pb and Cr - do not share any common properties, yet all of them induce superconductivity in FeTe. Moreover, although the quality (sharpness) of the superconducting transition varies depending on the type and amount of dopants, the superconducting onset temperatures are all very similar to one another, close to ~12 K. This suggests that the main role of the dopants is just to disturb the underlying AFM order of the FeTe system. Once the AFM order is disrupted by cationic impurities, the ground state of the FeTe system seems to switch from AFM to SC state, as sketched in the cartoon of Figure 4e. This picture is also in line with the long-standing theoretical proposal that FeTe is a special Hund metal located near an OSMP (orbital selective Mott phase), which is surrounded by many competing phases such as topological superconductivity, ferromagnetism, antiferromagnetism, stripe phases, and non-topological superconductivity.[32, 42] In this scenario, small perturbations, such as impurities, should be able to switch the ground state of FeTe from one to another, which seems to readily explain the emergence of SC state in the otherwise AFM ground state of FeTe, as depicted in Figure 4e. In other words, our findings suggest that interfacial cationic diffusion, or "chemical proximity effect," works as a perturbation to drive the unexpected yet universal superconductivity in various FeTe-based heterostructures. This observation also suggests that not only the chemical proximity effect but also other perturbations such as strain, magnetism, disorder, conducting interfacial states, etc. [22-24, 26, 27, [37] could also help transition the underlying AFM order of FeTe into the SC or other ground states. This will be an interesting line of future research directions. The bottom line is that FeTe heterostructures exhibit a unique bi-directional SC proximity effect in that once superconductivity is induced in the FeTe layer by its neighbor through the chemical (or other perturbative) proximity effect, the superconducting FeTe layer can proximitize the neighboring layer into a SC state through the regular proximity effect.

## 3. Conclusion

In conclusion, we have introduced the first Fe-based superconductor/ferromagnetic heterostructure, FTS/FGT, both of which are known to be correlated Hund metals. [32-35] The system becomes ferromagnetic below ~220 K and superconducting below ~13 K. MFM study shows that the FM state coexists with the SC state. Furthermore, we found that when non-superconducting FeTe is grown on top of the ferromagnetic FGT layer, the system becomes superconducting at a temperature comparable to that of the FTS/FGT system. We showed that Ge-interdiffusion is likely the main origin behind the emerging superconductivity in the FeTe/FGT system. Moreover, it is shown that superconductivity can be universally induced in the FeTe system with a wide range of other cationic impurities, including Bi, Mn, Pb, and Cr. Considering that such a minute level of interdiffusion is unavoidable in FeTe-based telluride heterostructures, superconductivity can universally emerge in FeTe heterostructures even if the component layers are not superconducting on their own. This also suggests that the ground state of FeTe is very close to the SC state such that it might be possible to drive it in and out of the SC state with a variety of other perturbations. Altogether, this study shows that Fe-Te-based heterostructures provide a unique opportunity to study a variety of novel physics and applications such as Hund metals, ferromagnetic superconductors, cryogenic spintronics, topological quantum computation, and Majorana physics, paving the way toward new superconducting technologies.

## 4. Experimental Section

*Thin-film growth*: We prepared FTS/FGT and FT/FGT films on 10×10×0.5 mm$^3$ Al$_2$O$_3$ (0001) and FT film on flexible mica (Ted Pella) substrates using a custom-built MBE system (SVTA) with a base pressure of ~10$^{-10}$ Torr. Al$_2$O$_3$ substrates were cleaned ex-situ by UV-generated ozone and in situ by heating to 750 °C under oxygen pressure of $5 \times 10^{-7}$ Torr. For the mica substrate, we exfoliate the contaminated surface right before loading it into the chamber and *in situ* cleaning at 400 °C for 15 min. High purities of Fe, Ge, Bi, Se, Mn, Pb, and Te were thermally evaporated using Knudsen diffusion cells for the film growth. All the source fluxes were calibrated *in-situ* by quartz crystal micro-balance and *ex-situ* by Rutherford backscattering spectroscopy. All capping layers were deposited at room temperature on top of the films.

*Transport measurements*: All transport measurements were performed using the standard Hall bar patterned samples with the channel length and width of 3 mm and 1 mm, respectively. Five indium wires were manually pressed on the Hall bar patterned sample used for four-probe longitudinal and Hall resistance measurements. Keithley 2400 source-measure units and 7001 switch matrix system were controlled by a LabView program for the longitudinal and Hall resistances measurements. Quantum anomalous Hall effect measurements were done in a Physical Property Measurement System (PPMS, Quantum Design inc.) down to 2 K.

*HAADF-STEM measurements*: High-angle annular dark-field scanning transmission electron microscopy (HAADF-STEM) samples were prepared by a focused ion beam (Helios G5, Thermo Fischer Scientific) using 2.0 keV Ga$^+$ ions for final milling. HAADF STEM images were acquired with 200 keV electrons with collection angles ranging from 67 to 275 mrad using double aberration-corrected JEOL.

*Magnetic force microscopy (MFM)*: MFM images of Fe(Te,Se)/Fe$_3$GeTe$_2$ were taken by using a homemade Helium-3 MFM system equipped with *in situ* transport measurement capability. Commercial piezoresistive cantilevers with a spring constant k of ~3 N/m and a resonant frequency $f_0$ of ~44 kHz were employed, and the tips were coated with ~100-nm-thick Co using magnetron sputtering. MFM imaging was conducted in constant-height noncontact mode. The MFM signal (the resonant frequency shift) is proportional to the out-of-plane magnetic force gradient, which was extracted by a phase-locked loop (SPECS).

## Supplementary information

Supporting Information is available from the Wiley Online Library or from the author.

## Acknowledgements

This work is supported by Army Research Office's W911NF2010108, MURI W911NF2020166, and the center for Quantum Materials Synthesis (cQMS), funded by the Gordon and Betty Moore Foundation's EPiQS initiative through Grant GBMF10104. X.Y. is supported by the National Natural Science Foundation of China (Grant No. 12304541). The MFM work on FTS/FGT film is supported by the Office of Basic Energy Sciences, Division of Materials Sciences and Engineering, and US Department of Energy under Award No. DE-SC0018153. The scanning transmission electron microscopy work performed at Brookhaven National Laboratory is sponsored by the US Department of Energy, Basic Energy Sciences, Materials Sciences and Engineering Division, under contract no. DE-SC0012704. This research used the Electron Microscopy resources (the Helios G5 FIB) of the Center for Functional Nanomaterials (CFN), which is a U.S. Department of Energy Office of Science User Facility, at Brookhaven National Laboratory under Contract No. DE-SC0012704. The XRD work performed at Oak Ridge National Laboratory. We would like to thank G. Kotliar, P. Coleman, J. Pixley and G. Blumberg at Rutgers University for helpful discussions.

## Author contributions

S.O. and H.T.Y. conceived the experiments. H.T.Y., X.Y., and D.J. performed thin film growth and H.T.Y. performed transport measurements. Y.-T.C and K.D. performed MFM. M.H., K.K., and Y.Z. performed the STEM. M. B and A.-S.C. performed XRD. S.O., H.T.Y., W.W, and S.-W.C. analyzed the data. S.O. and H.T.Y. wrote the paper with feedback from other coauthors.

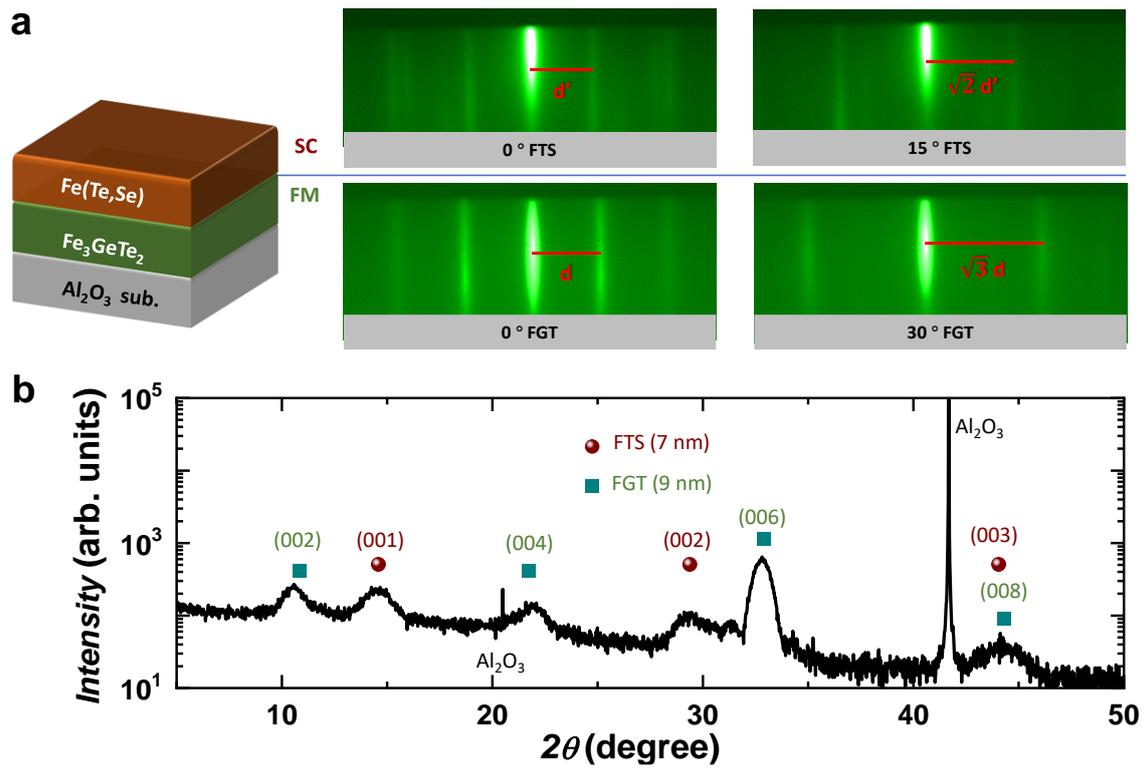

**Figure 1.** Structural analysis of an FTS/FGT heterostructure. a) Schematic of the SC/FM heterostructure and reflection high-energy electron diffraction (RHEED) patterns at two high-symmetry directions of 7 nm $FeTe_{6/7}Se_{1/7}$ (top panels) and 9 nm $Fe_3GeTe_2$ (bottom panels) film grown on $Al_2O_3(0001)$ substrate. b) X-ray diffraction data for the FTS/FGT heterostructure. Circle and square symbols indicate the peaks for FTS and FGT, respectively.

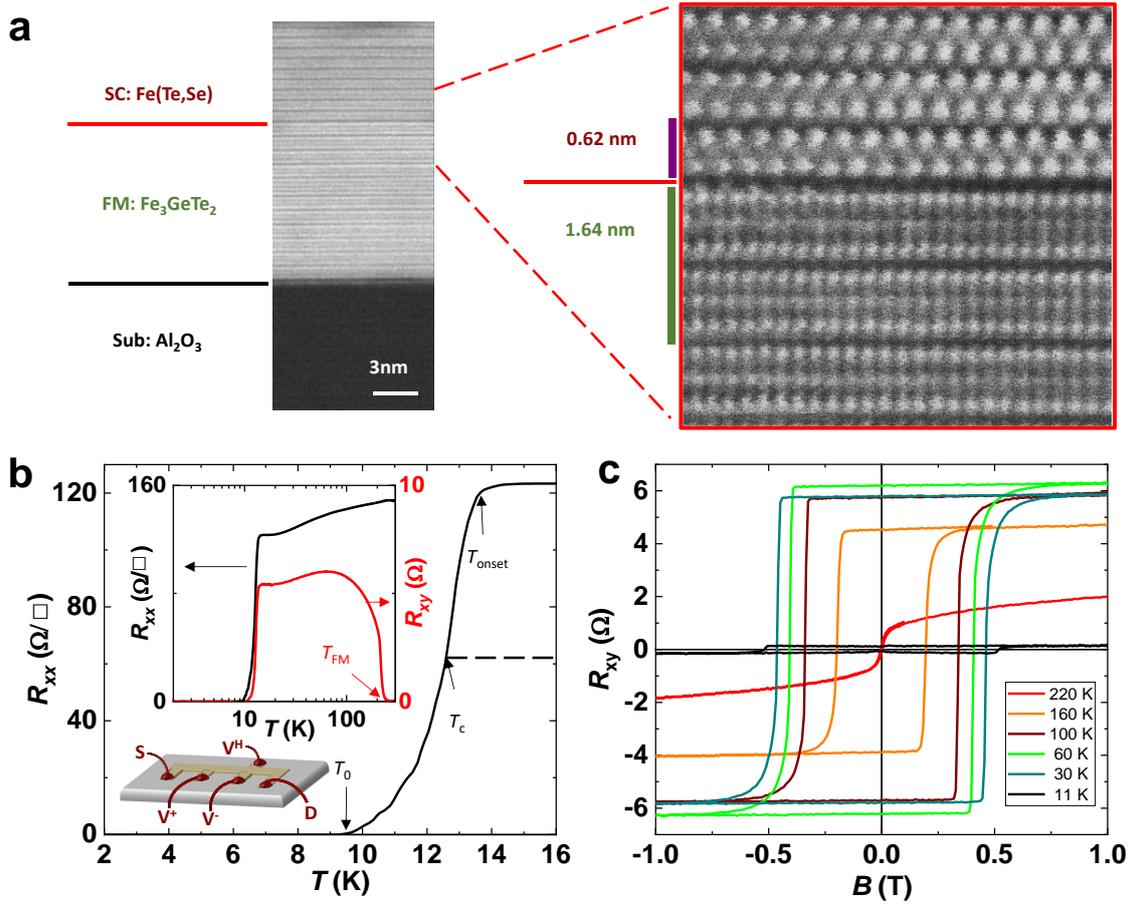

**Figure 2.** STEM image and transport properties of FTS/FGT film. a) High-angle annular dark-field scanning transmission electron microscopy (HAADF-STEM) image of an FTS/FGT sample. b) Temperature-dependent longitudinal sheet resistance, $R_{xx}$, of FTS/FGT, and a semi-log plot of $R_{xx}$ (black) and $R_{xy}$ (red) from 300 K to 2 K in the inset, recorded without an external magnetic field. The dashed line indicates half the value of the normal state resistance. The schematic shows the Hall bar patterned sample with a channel length of 3 mm and width of 1 mm. c) Field-dependent Hall resistance at various temperatures: 11 K, 30 K, 60 K, 100 K, 160 K, and 220 K.

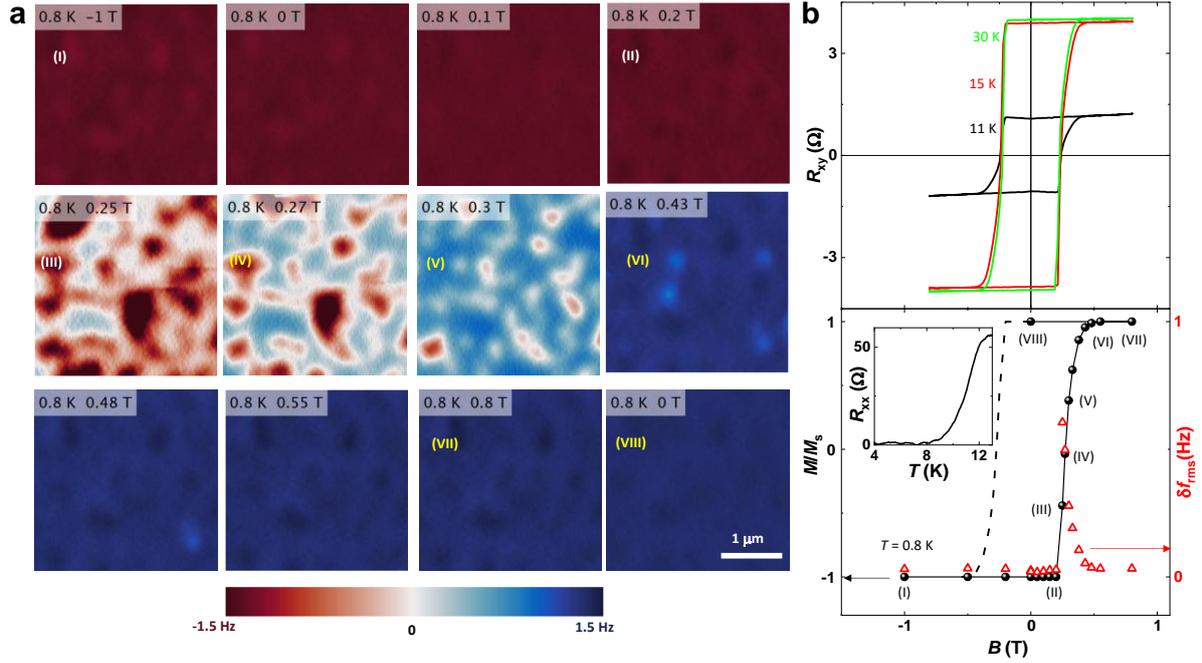

**Figure 3.** MFM images of FTS/FGT film in the superconducting state and comparison with the anomalous Hall effect data in the normal state. a) Field-dependent MFM images of FTS/FGT at 0.8 K, well below $T_c$. Sequential MFM images from (I) -1 T (saturation state with negative field), (II) 0.2T, (III) 0.25 T (the coercivity field), (IV) 0.27 T, (V) 0.3 T, (VI) 0.43, (VII) 0.8 T (saturation state with positive field), and (VII) 0 T (remanence state). b) Top panel: Hall resistance as a function of an external magnetic field at various temperatures, measured using a PPMS (Quantum Design) just before transfer to the MFM system. Bottom panel: the normalized magnetization $M/M_s$ (black circles) and $\delta f_{rms}$ (red triangles) as a function of applied magnetic fields obtained from MFM images, exhibiting typical ferromagnetic hysteretic behavior. The inset shows the in situ temperature dependence of the two-probe resistance, which was measured using the He-3 MFM system. The resistance (after removing contact resistance) curve clearly shows a superconducting phase transition around 10-12 K. This *in situ* transport measurement confirms that superconductivity coexists with ferromagnetism.

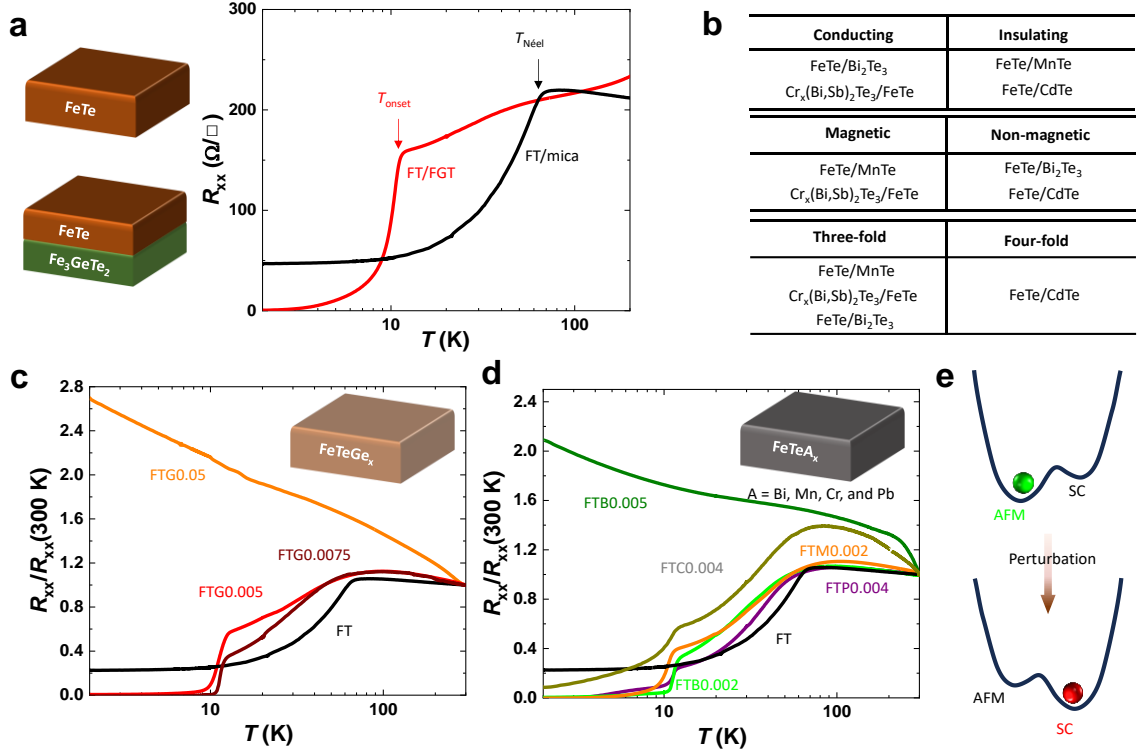

**Figure 4.** Temperature-dependent resistance of various control samples. a) Semi-log plot of $R_{xx}$ versus temperature characteristics for FeTe/mica (black) and FeTe/FGT (red), respectively. b) Comparison of magnetism, conductivity, and structural symmetry of neighboring tellurides that exhibit superconductivity when used in FeTe heterostructures. Temperature-dependent $R_{xx}$ of doped FeTe films grown on mica substrates. c) 5 % Ge-doped (FTG0.05), 0.75 % doped (FTG0.0075), and 0.5 % doped (FTG0.005) FeTe films. d) 0.5 % Bi-doped (FTB0.005), 0.2 % Bi-doped (FTB0.002), 0.4 % Cr-doped (FTC0.004), 0.4 % Pb-doped (FTP0.004), and 0.2 % Mn-doped (FTM0.002) FeTe films. $R_{xx}(T)$ data (black) of FeTe/mica is included for comparison in both c) and d). e) Cartoon describing the shift of the ground state from AFM to SC state in the FeTe system with external stimuli such as doping. Note that despite varying qualities of the SC transitions, the SC onset temperatures are all very similar at ~12 K, regardless of the doping details. This strongly suggests that the main role of dopants is just to shift the ground state of the FeTe system from AFM to SC state as sketched in e).

Supporting Information

**Universal superconductivity in FeTe and all-iron-based ferromagnetic superconductor heterostructures**

*Hee Taek Y\*, Xiong Yao, Deepti Jain, Ying-Ting Chan, An-Hsi Chen, Matthew Brahlek, Kim Kisslinger, Kai Du, Myung-Geun Han, Yimei Zhu, Weida Wu, Sang-Wook Cheong, and Seongshik Oh\**

## Temperature-dependent $R_{xx}$ of the FTS/FGT system under varying magnetic fields

We analyze the superconducting transport properties of the FTS/FGT system. Figure $S$1a and $S$2b show the temperature-dependent normalized longitudinal resistance $R_{xx}/R_n(T,B)$ under external magnetic fields along perpendicular and parallel directions to the film surface. As we increased the magnetic fields from 0 to 9 T along $B//c$, we observed a significant broadening of $T_c$. On the other hand, in the $B//ab$ direction, the reduction in $T_c$ was relatively minimal. Current ($I$) vs voltage ($V$) characteristics near $T_c$ in Figure $S$1c and d show that the superconducting behavior is well described by the Berezinskii-Kosterlitz-Thouless transition (BKT) of vortex-antivortex pairs in 2d. [1-3] We show in Figure $S$1c, a temperature-dependent, log-log $I$-$V$ plot, with $V \propto I$ (black dashed line) and $V \propto I^3$ (red dashed line) to compare with the power-law equation of $V \propto I^{\alpha(T)}$. The detailed temperature dependence of the exponent ($\alpha$) in Figure $S$1d provides superconducting onset temperature of 13.5 K with $\alpha = 1$, and BKT temperature of 12.2 K with $\alpha = 3$, both of which are well matched with the $T_{\text{onset}}$ of 13.5 K and the $T_c$ of 12.5 K observed in the $R$-$T$ data of Figure 2b. A close match of $T_{\text{BKT}}$ and $T_c$ indicates that this system follows the BKT transition.

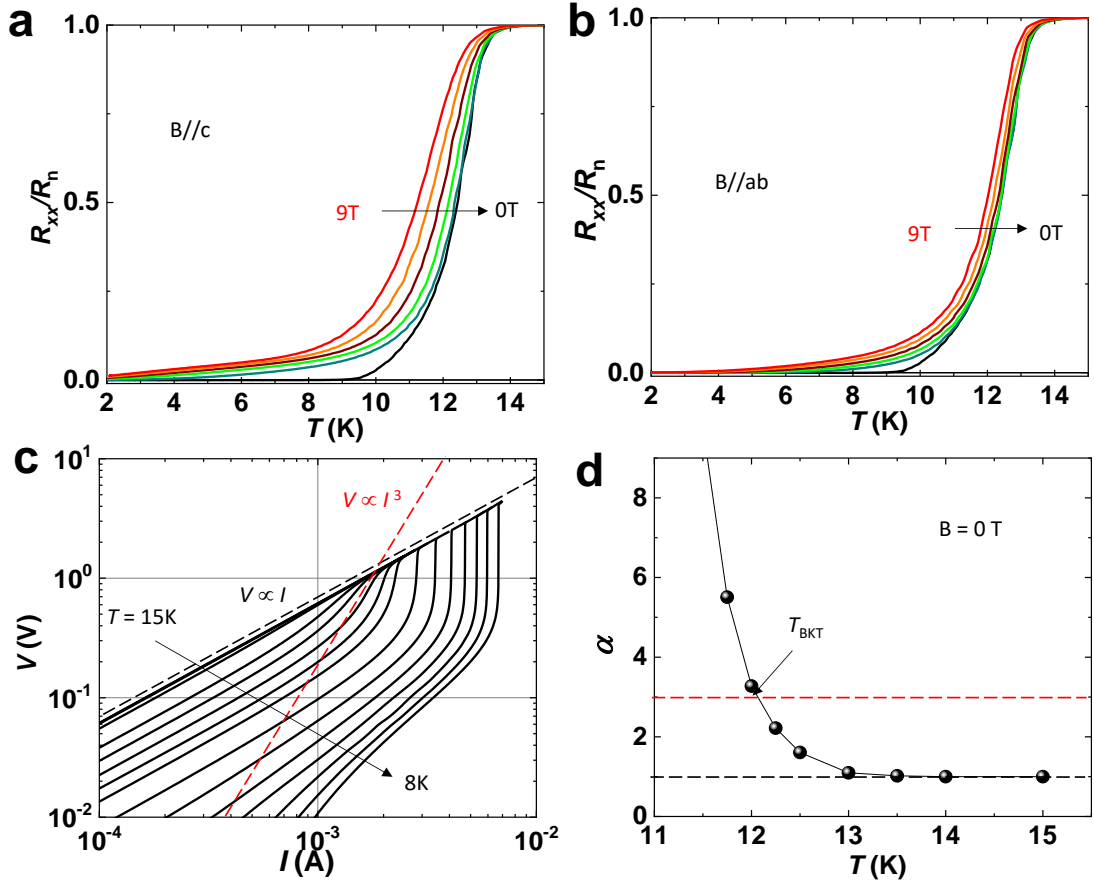

**Figure $S$1.** Temperature-dependent $R_{xx}$ of the FTS/FGT heterostructure under varying magnetic fields from 0 to 9 T along. a) perpendicular b) parallel to the ab plane. c) Double-logarithmic, voltage $V$ versus current $I$ plot for temperatures near $T_c$. The black and red dotted lines indicate the $\alpha$ value of 1 and 3, respectively, in the power-law equation of $V \propto I^{\alpha(T)}$. d) Temperature-dependent exponent $\alpha$ near $T_c$. $T_{\text{BKT}}$ is determined by the temperature at which $\alpha = 3$.

*Magnetic force microscopy images of Fe(Te,Se)/Fe$_3$GeTe$_2$ system*

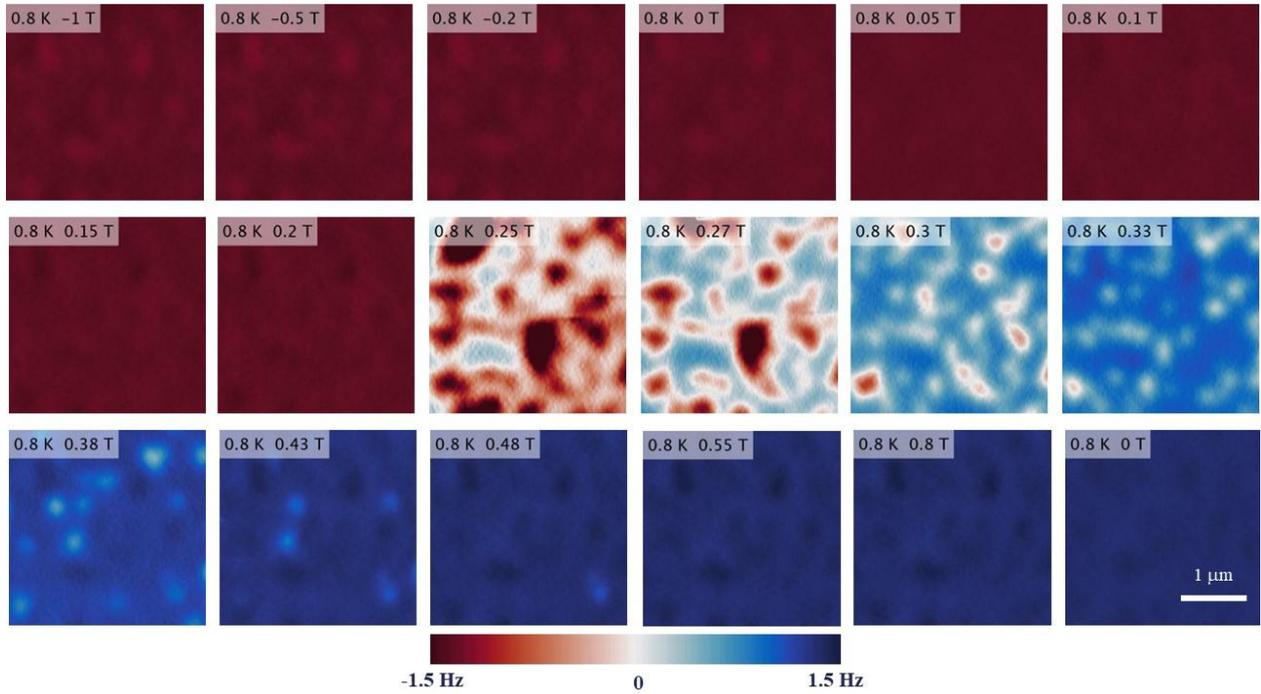

**Figure *S*2.** Complete data set of field-dependent MFM images of Fe(Te,Se)/Fe$_3$GeTe$_2$ below the superconducting transition temperature. The MFM images are taken at 0.8 K. The varying magnetic field is shown at the top left corner of each MFM image. The lift height is 80 nm. The tiny changes between 0 and 0.1 T are caused by the reversal of the tip moment due to the external magnetic field.

*Hall resistance of FeTe/Fe$_3$GeTe$_2$ system*

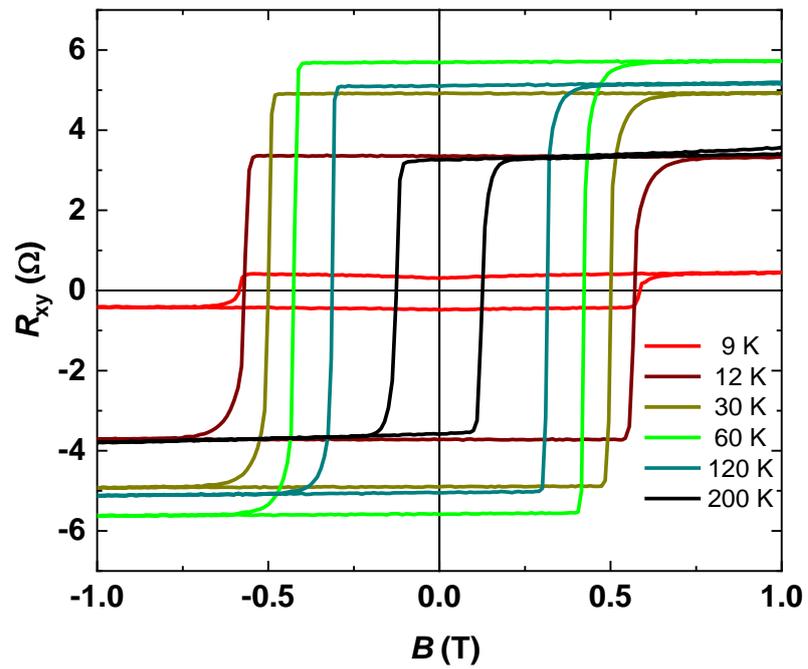

**Figure *S*3.** Field-dependent Hall resistance of the FeTe/Fe$_3$GeTe$_2$ heterostructure at various temperatures: 9 K, 12 K, 30 K, 60 K, 120 K, and 200 K.

*Temperature-dependent $R_{xx}$ of the doped FeTe thin films*

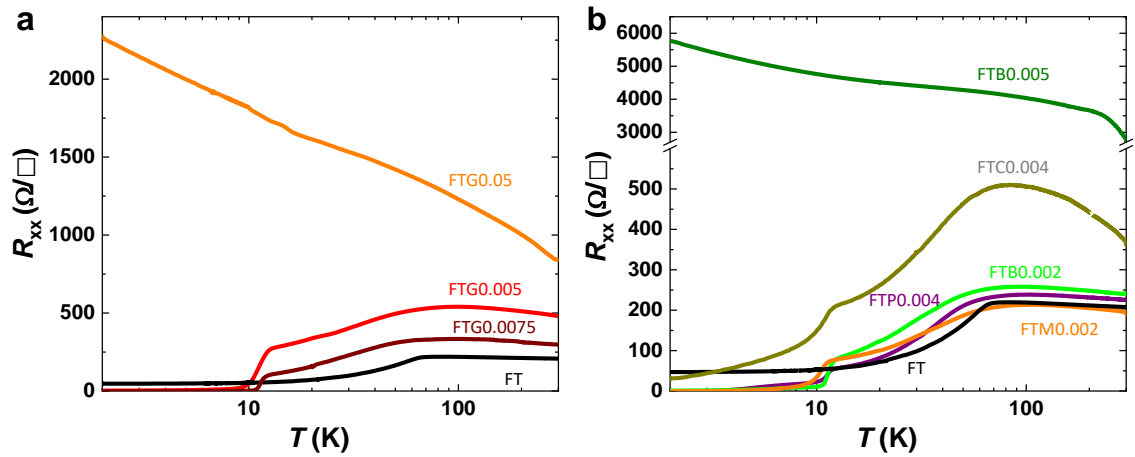

**Figure *S*4.** Temperature-dependent $R_{xx}$ of the doped FeTe thin films. a) 5 % Ge-doped (FTG0.05), 0.75 % doped (FTG0.0075), and 0.5 % doped (FTG0.005) FeTe films. $R_{xx}(T)$ data (black) of FeTe/mica is included for comparison. b) 0.5 % Bi-doped (FTB0.005), 0.2 % Bi-doped (FTB0.002), 0.4 % Cr-doped (FTC0.004), 0.4 % Pb-doped (FTP0.004), and 0.2 % Mn-doped (FTM0.002) FeTe films. $R_{xx}(T)$ data (black) of FeTe/mica is included for comparison in both a) and b).